\documentclass[conference, draftclsnofoot, onecolumn]{IEEEtran}
\usepackage{graphicx}
\usepackage[nocompress]{cite}
\usepackage{cite}
\usepackage{amsmath}
\usepackage{subcaption}

\begin{document}
	\title{A Resources Representation For Resource Allocation In Fog Computing Networks}
	
	\author{\IEEEauthorblockN{Amine Abouaomar$^{*1,2}$, Soumaya Cherkaoui$^{*1}$, Abdellatif Kobbane$^{*2}$, Oussama Abderrahmane Dambri$^{*1}$
		}
		\IEEEauthorblockA{
			$^1$ INTERLAB, Engineering Faculty, Universit\'e de Sherbrooke, Canada. \\
			$^2$ Rabat IT Center, ENSIAS, Mohammed V University of Rabat, Morocco.}
	}

	\IEEEtitleabstractindextext{%
		\begin{abstract}
			Fog computing is emerging as a new paradigm to deal with latency-sensitive applications, by making data processing and analysis close to their source. Due to the heterogeneity of devices in the fog, it is important to devise novel solutions which take into account the diverse physical resources available in each device to efficiently and dynamically distribute the processing. In this paper, we propose a resource representation scheme which allows exposing the resources of each device through Mobile Edge Computing Application Programming Interfaces (MEC APIs) in order to optimize resource allocation by the supervising entity in the fog. Then, we formulate the resource allocation problem as a Lyapunov optimization and we discuss the impact of our proposed approach on latency. Simulation results show that our proposed approach can minimize latency and improve the performance of the system.
		\end{abstract}
		
		\begin{IEEEkeywords}
			Edge Computing, Resources Representation Model, Resource allocation, Lyapunov optimization.
	\end{IEEEkeywords}}

	\maketitle
	
	\IEEEdisplaynontitleabstractindextext
	\IEEEpeerreviewmaketitle

	\section{Introduction}\label{sec:introduction}
	In the last few years, fog computing attracted the attention of both academia and industrial. Fog computing is an emerging paradigm consisting in extending cloud-based computing, storage and networking capabilities to the edge of the network \cite{fog-computing, Rachedi}. Fog computing main purpose is reducing the latency and ensuring an optimized use of available resources at the edge of the network. Therefore, resources of fogs need to be managed and used appropriately.
	
	The fog is built-up on fog nodes \cite{fog-node, abouaomar1, abouaomar2}. Each fog node is an aggregation of edge devices such as routers, switches, and edge devices \cite{Azizian2, Azizian1}. From the perspective of location, the devices forming fog nodes are, usually, outside the premises of the operator, which makes their management and maintenance difficult. From the architectural perspective, these devices may have different designs and diverse capabilities. Also, there can be different types of devices in the fog. We distinguish for example, routers, servers, switches \cite{edn}. Consequently, some edge devices may have specialized or powerful processing resources such as Graphical Processing Units (GPUs), which make them able to perform complex and heavy computational tasks. Other devices may have big storage space, making them more suitable for caching and storing at the edge of the network. Finally, devices such as routers and switches, with moderate processing but with powerful networking capabilities can handle more traffic than others. 
	
	The unavailability of information about devices' capabilities at the fog nodes might cause an unnecessary delay; because some devices might be required to perform tasks they are not the best suited for performing. Therefore, exposing the resources of fog devices to the supervising entity will help distributing the fog tasks in an optimal manner. Indeed, the supervising entity can build adequate resource allocation schemes from these information.
	
	Fog computing generally relies on virtualization because of its capacity to represent heterogeneous systems. The concept of virtualization allows running an entire Operating System (OS) with its applications on a virtual machine (VM). However, VMs need a significant startup time to run on hardware platforms, which may cause an unnecessary delay for resources consumption in fog nodes. This delay can be avoided by using containers. Containers have been proposed as a lightweight virtualization platform, which offers much more advantages than those offered by traditional virtualization.  Containers have a lower response time since their startup time is much smaller than VMs (Milliseconds VS Seconds). Containers are superior to VMs in terms of CPU, memory, and disk performance. Also, VMs' network performance in term of response time is lower than that of containers (microseconds) \cite{rescont}.
	
	In this paper, we propose a resource representation model, and we test the ability to discover and represent resource information independently of the operating system, the architecture or the manufacturer. In this work, resources of interest are processing, storage, memory and networking capabilities. We also formulate the resource allocation optimization problem aiming at minimizing the system latency, and we use a Lyapunov model to solve it. In order to evaluate the performance of the  proposed approach, we perform simulations for two scenarios with Docker containers \cite{docker}, a scenario with resource representation and a scenario without resource representation.
	
	The main contributions of this paper are summarized as follows:
	\begin{itemize}
		\item We propose a resource representation model for physical resources in a fog environment which adheres to  European Telecommunications Standards Institute (ETSI) standard ETSI-GS-MEC-009. We use this model to represent various types of resources (processors, memory, storage, and networking) of fog nodes and their capacities.
		\item We propose a resource allocation model based on Lyapunov optimization to minimize the experienced delay. Additionally, we use containers as a lightweight virtualization mechanism to simulate our proposed method.
	\end{itemize}
	
	To the best of our knowledge, this is the first paper to study resource representation and its usage to optimize resource allocation at the fog level. The rest of this paper is organized as follows: we presents the system model in Section II. Section III is dedicated to the discussion of the proposed approach. Section IV details the simulations and the corresponding results. Related works are discussed in Section V. Finally we conclude the paper in Section VI.
	\section{System Model}
	\begin{figure}[h]
		\centering
		\includegraphics[width=.7\linewidth]{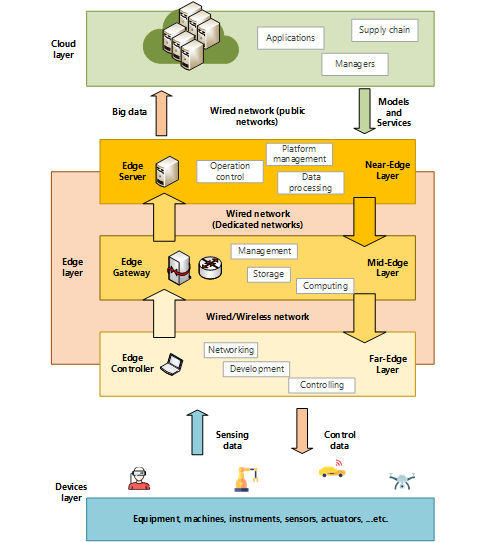}
		\caption{Generic Fog Computing architecture.}
		\label{fig:fog-scheme}
	\end{figure}
	The fog computing architectural deployment consists of three layers, as depicted in Fig. \ref{fig:fog-scheme}. At the bottom, the end users, we distinguish two types of users. The first type is the data producers; all kind of devices capable of intercepting data about the environment and transform it into an information (sensors, actuators, phones, ...etc). The second type of users is the data consumers such as mobile devices, applications or even sensors.\\
	The second layer is the fog layer, where devices are aggregated to form fog nodes. These devices could be servers, computers, mobile devices or edge routers and switches. This layer is dedicated to perform the data pre-processing and real-time analytics. It is the perfect placement for the latency-sensitive application to do processing. It could serve as caching infrastructure in caching use cases.\\
	Finally, on the top, the cloud and data centers layer, devoted to long-term analytics (latency non-sensitive applications) and for permanent storage use cases.

	Let $\mathcal{U}$ be the set of users.
	\[
	\mathcal{U} = \{u_1, u_2, ..., u_N\}
	\] 
	distributed over a geographical area, covered by fog nodes. Let $\mathcal{FN}$ be the set of fog nodes. 
	\[
	\mathcal{FN} = \{fn_1, fn_2, ..., fn_M\}
	\]
	Each fog node serves a sub set of $\mathcal{U}$. A fog node is formed with $K$ devices, with $K$ is variable depending on which fog.
	\[
	\mathcal{D}_{i} = \{d_{(i,1)}, d_{(i, 2)}, ..., d_{(i, K)}\}
	\]
	Each device has a set of resources denoted $\mathcal{R}_{(i, j)}$.
	\[
	\mathcal{R}_{(i, j)} = \{r_{(i, j)}^1, r_{(i, j)}^2, ..., r_{(i, j)}^P\}
	\]
	where $r_{(i, j)}^k$ represent the $k^{th}$ resource  of the $j^{th}$ device in the $i^{th}$ fog node.\\
	where $r_{(i, j)}^k$ $\in$ $\{Processing, Storage, Networking\}$, and $P$ is the number of resources.\\
	Each device host a number $Q$ of containers denoted by $\mathcal{C}_{(i, j)}$, in the $i^{th}$ fog node at the $j^{th}$ device. $Q$ the number of containers at each device differs from the others.
	\[
	\mathcal{C}_{(i, j)} = \{c_{(i, j)}^1, c_{(i, j)}^2, ..., c_{(i, j)}^Q\}
	\]

	\section{Proposed Approach}
	In this section we will present the proposed resources representation model and how the entities communicate.
	\subsection{Unified Resources Representation}
	In this paper we are interested in the physical resources (processing unites, memory, storage and networking) of the devices forming the fog nodes. We distinguish three types of resources. Resources dedicated to processing, including the CPU, GPU and other computing units plus the memory. The other type of resources is devoted to storing such as hard disks (HDD) and solid state disks (SSD). Finally, the networking resources.
	\begin{figure}
		\centering
		\includegraphics[width=0.5\linewidth]{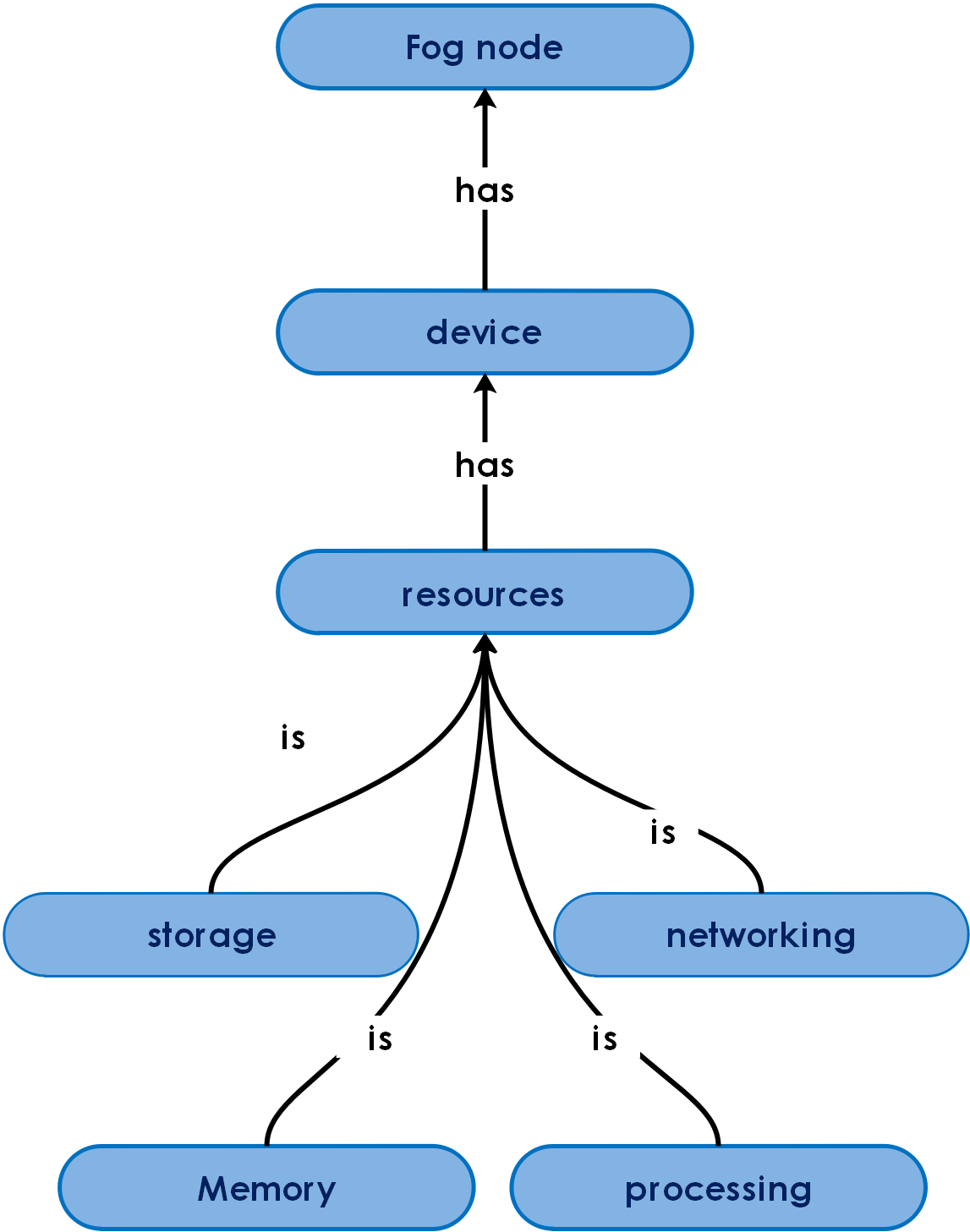}
		\caption{An overview of the resources families from a general point of view.}
		\label{fig:globalenew}
	\end{figure}
	
	To discover these resources, we first need to get information about it. To do so, we used low-level operations and commands such as \textit{CPUID} to get the CPU properties such as the architecture, number of cores and the frequency. We also used commands such \textit{cpuinfo} and \textit{meminfo} for Linux based machines to get the current state of the CPU usage and memory respectively. For the Windows based devices, \textit{wmic} command provide the same information. And so on for the majority of the used OS in the devices forming the edge layer.
	
\begin{figure}[h]
		\centering
		\includegraphics[width=.7\linewidth]{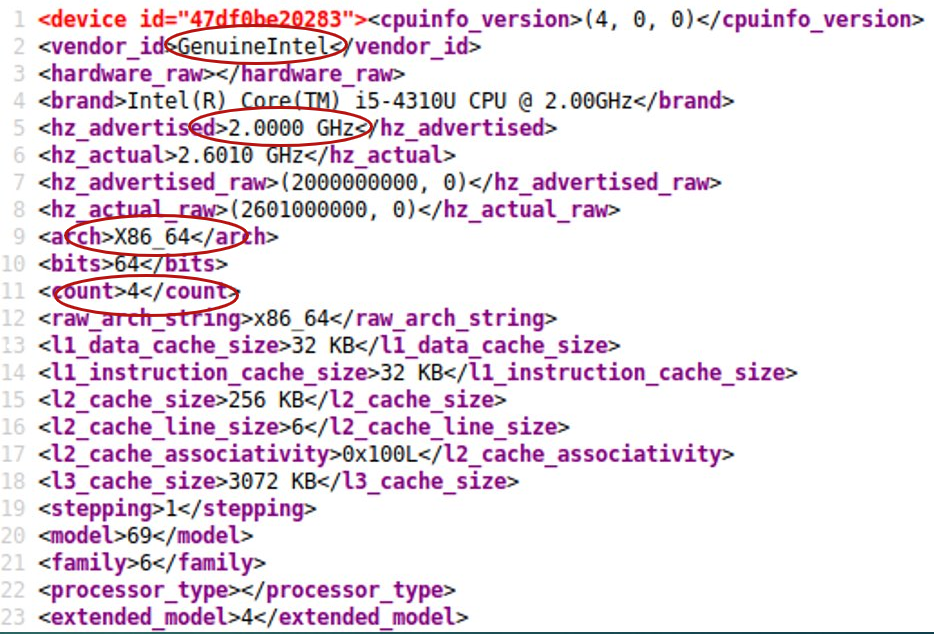}
		\caption{Example of a response generated by a device being requested for CPU information.}
		\label{fig:exmpl}
	\end{figure}

	Gathering all these commands within the same program that, depending on the Operating System used, it get the information about the devices and their available resources. Fig. \ref{fig:exmpl} shows an example of a device response when the supervisor requested information about the CPU capabilities. The output is an XML response with all the different information about the CPU such as the family, the frequency, number of cores and the architecture.
	
	As described in the standard IETF RFC 3986 \cite{std:3986} and the ETSI MEC APIs ETSI-GS-MEC-009 described in \cite{edge-etsi} for developing edge computing APIs. The supervising entity uses the program to get the resources availability of the devices forming the fog node. It performs a request over the nodes and get the whole information or just the needed ones. The nodes expose their resources in XML format. After that the supervisor decide what strategy to adopt in the process distributing.

	\begin{figure}[h]
		\centering
		\includegraphics[width=\linewidth]{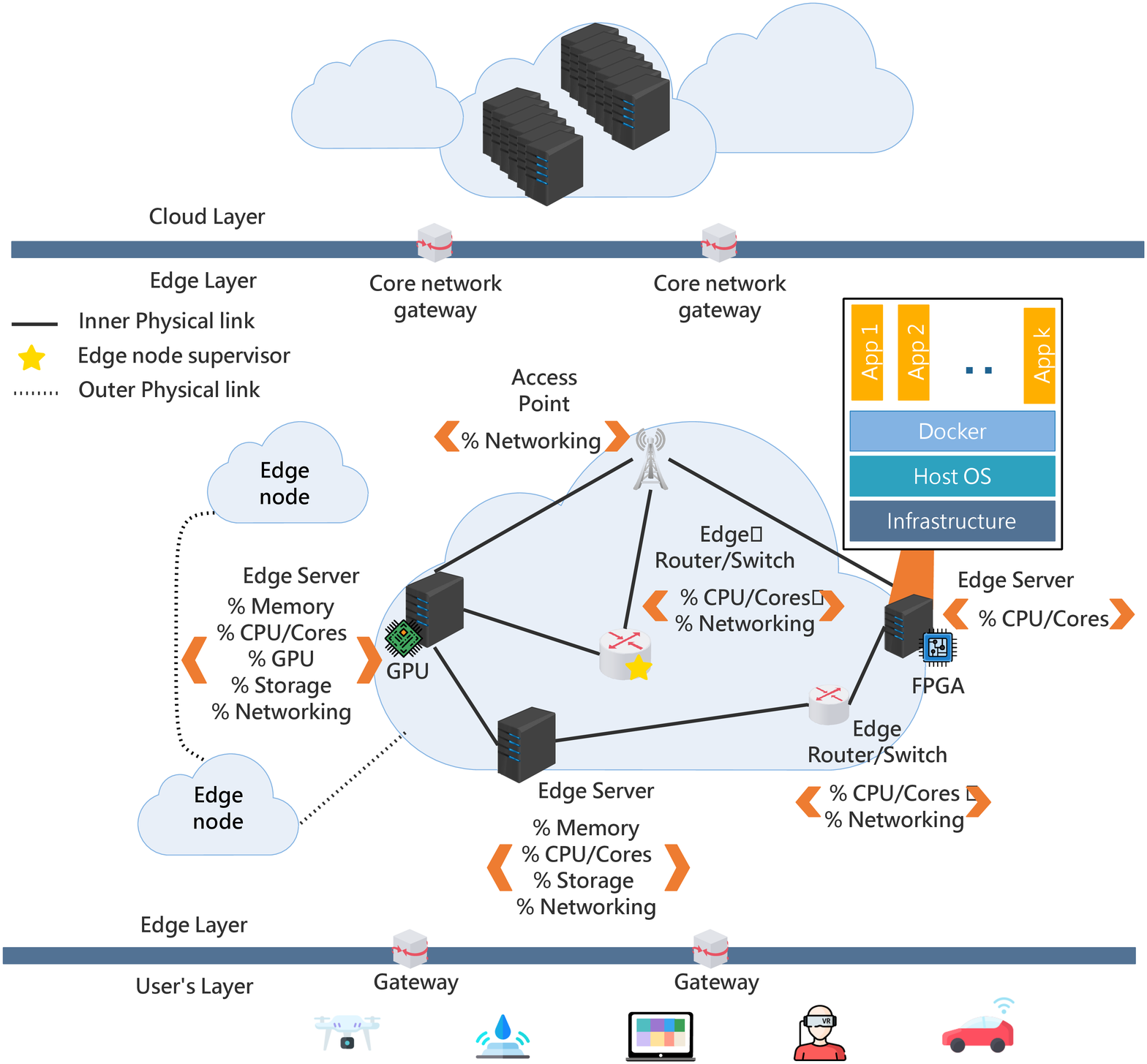}
		\caption{Overview of the system model}
		\label{fig:arch-lyapunov}
	\end{figure}
	
	As illustrated in Fig. \ref{fig:arch-lyapunov}, users depicted at the bottom side of the figure request tasks from the fog environment (ex. processing tasks to render a virtual reality environment). The requests will be forwarded to the fog nodes’ supervising entity. The devices in the fog expose their resources through ETSI MEC APIs to the supervising entity which, in turn, adequately distributes the requested task to the suitable containers based on the information gathered from the fog devices.

	\subsection{Resources Allocation}
	For each container, a portion of the hosting device's resources is allocated. Let $\gamma_{r_{(i, j)}^k}$ be a binary variable to denote if the resource $r_{(i, j)}^k$ is allocated to the container. And let $\theta_{r_{(i, j)}^k}$ be the amount of resource $r_{(i, j)}^k$ allocated to the container $c_{i, j}^l$. We can express the amount of resources allocated to all the containers at the instant $t$ as follows : 
	\begin{equation}
	Allocated(c_{(i, j)}, t) = \sum_{o \in R_{(i, j)}} \gamma_{r_{(i, j)}^o} . \theta_{r_{(i, j)}^o}
	\label{eq:alloc}
	\end{equation} 
	where $\gamma_{r_{(i, j)}^k} = 1$ if the $r_{(i, j)}^k$ is allocated, and equal to $0$ otherwise.
	The overall load on a device $j$ is the difference between its maximum capacity and the sum of all containers allocated resources : 
	\begin{equation}
	Load(d_{(i, j)}, t) = max(r_{(i, j)}) - \sum_{c \in \mathcal{C}_{(i,j)}} Allocated(c, t)
	\end{equation}
	Considering that the communication links are reliable, the experienced delay at a given node, depend essentially on the size of the task and the size of the queue at the container of the requested service. Let $\lambda_{(i,j)}^{k}$ be the mean of a Poisson Process of the requests arrival at the $i^{th}$ fog and the $j^{th}$ device for the container $k$. The experience delay at the instant $t$ is expressed as follows:
	\begin{equation}
	Delay_{(i, j)}(t) = \frac{\lambda_{(i,j)}^{k}}{C_{i, j}^{max}} + Queue_{(i, j)}^k(t)
	\label{eq:queue_cont}
	\end{equation}
	where $C_{i, j}^{max}$ is the capacity of resources allocated to the container $k$. And $Queue_{(i, j)}^k(t)$ is the queue size of the same container.
	Let the set $Req_i$ be the set of requests to be processed by the fog node $fn_i$. Inspired from \cite{omar-chakroun}, we formulated the queue dynamic for a given container $c_{(i, j)^k}$ hosted at a device $j$ as follows : 
	\begin{equation}
	\omega_{(i, j)}^k(t+1) = \omega_{(i, j)}^k(t) - Rq_{(i, j)}(t) + \sum_{Requests} Rq_{(i,j)}^k(t)
	\end{equation}
	where $Rq_{(i, j)}^k(t)$ and $Rq_{(i, j)}(t)$ are the requests for the container $k$ at the device $d_{(i, j)}$ of the fog node $fn_i$ and the total requests to be processed by the fog node $fn_i$ at the instant $t$.
	In order to distribute, the supervisor of the fog node needs to adopt a distribution policy first, this distributed policy depends essentially on the exposed resources by the devices that the supervising entity consult in real time to get informed.\\
	Let $\pi_i$ be the distribution policy adopted by the supervisor to be applied to a fog node $fn_i$. The queuing dynamic of the fog node $fn_i$ is given by :
	\begin{equation}
	U_{(i, j)}(t+1) = max (U_{(i, j)}(t) - \mu_{(i, j)}(\pi_i(t)), 0) + Rq_{ij}(t)
	\label{eq:qeueue-fog}
	\end{equation}
	
	Where $\mu_{(i, j)}(\pi_i(t))$ is the demand rate for a service handled by a container hosted in a device $d_{(i, j)}$. In \cite{omar-chakroun} they proposed that the service rate could be derived from the processor frequency assignment. In our case, the resources representation service is used to get the information about the different resources as discussed in the previous section.
	
	For each container, $c_{(i, j)^k}$ the supervisor chooses the number of the requests to be handled by the device $d_{(i, j)}$. This number is the solution of the following problem : 
	
	\begin{subequations}
		\begin{equation}
		\label{eq:obj1}
		Maximize~:~Rq_{(i, j)(t)}[V\alpha_{(i, j)} - \omega_{(i, j)}(t)]
		\end{equation}
		\begin{equation}
		subject~to~:~Rq_{(i, j)}(t) \leq Rq_{(i, j)}^{new}(t)
		\end{equation}
	\end{subequations}
	$V$ is a non-negative control parameter chosen by the supervisor to specify the rate of a trade-off between the performance and the delay. $\alpha_{(i, j)}$ is the weight of the accepted requests average demand, $U_{(i, j)}(t)$ and $\omega_{(i, j)}(t)$ are evolving in time and represent, the queue backlog of the fog node and the queue of the container respectively.
	The solution for the problem in (\ref{eq:obj1}) could be solved by fixing thresholds. We suppose for example that the requests are admitted if :
	\[
	\omega_{(i, j)}(t) > V.\alpha_{(i, j)}
	\]
	and are not admitted if we have : 
	\[
	\omega_{(i, j)}(t) \leq V.\alpha_{(i, j)}
	\]
	In order to allocate the resources, let $\pi_i(t)$ the policy that represent the solution of the following optimization problem.
	\begin{subequations}
		\begin{equation}
		Minimize~:~\lim\limits_{t\to\infty} \frac{1}{t}\sum_{\tau=0}^{t-1}\mathbf{E}\left[Delay_{(i, j)}(\tau)\right]U_{(i, j)}(t)
		\end{equation}
		\begin{equation}
		subject~to~:~Delay_{(i, j)}(t) \leq D^{max} 
		\end{equation}
		\label{eq:opt-prob}
	\end{subequations}
	
	Where $D^{max}$ is the maximum delay experienced by the users when requesting a service offered by a container hosted at the device $d_(i, j)$ at the peak period, it is also the maximum experienced delay when the demand rate is for a given service is high.
	
	\section{Simulations and Results}

	\begin{figure}[h]
		\centering
		\includegraphics[width=.7\linewidth]{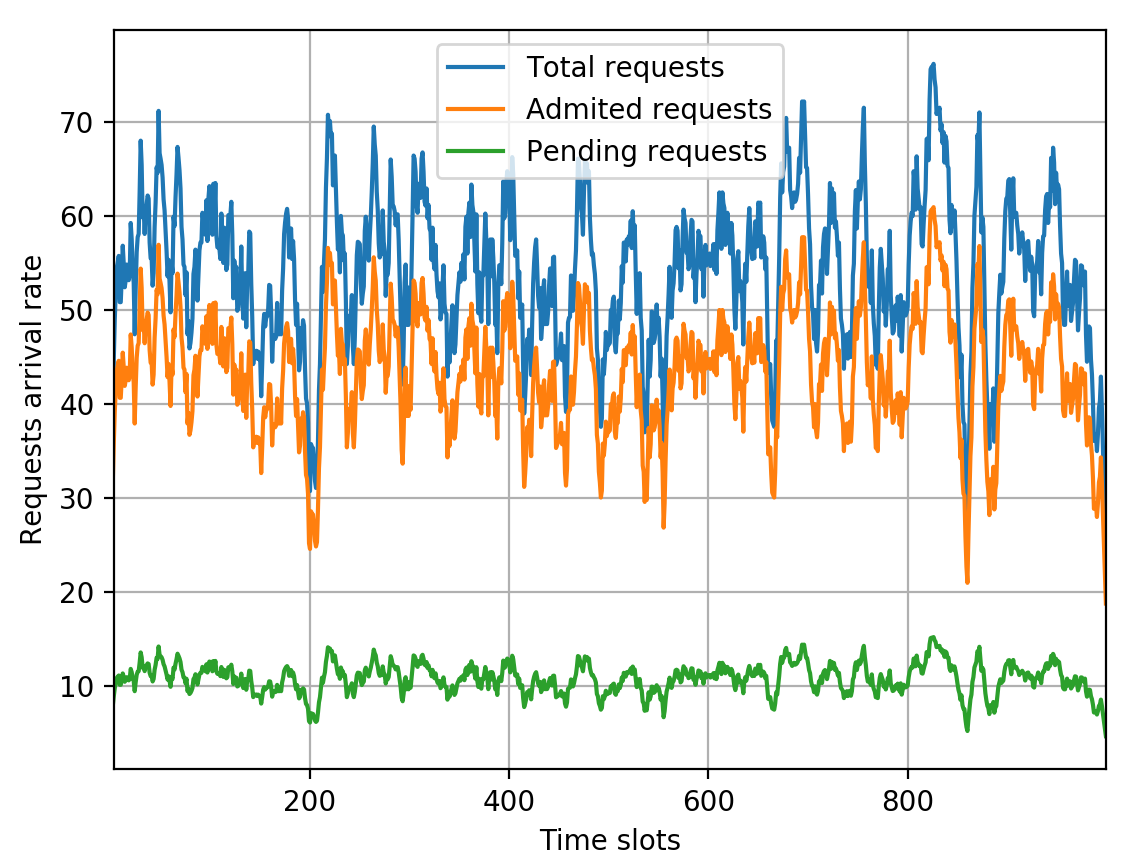}
		\caption{The average requests arrival, the admitted requests and requests on pending.}
		\label{fig:figure2f}
	\end{figure}

	We simulated a network of $10$ fog nodes. Each fog is formed by a number of devices varying between $5$ and $50$, where each device is hosting from $10$ to $100$ containers. We consider that the requests arrive at a rate within $0$ and $100$ per time slot. The capacity of processing vary between $5$ to $10$ request per time slot. The simulation duration is $500$ time slot. We simulated the system in $3$ cases of the $V$ parameter configuration $(0.2, 0.5, 0.9)$.
	
	Fig. \ref{fig:figure2f} illustrate the requests arrival with the accepted, admitted and pending requests, respectively on the top, middle and bottom.
	
	\begin{figure}[b]
		\centering
		\includegraphics[width=.7\linewidth]{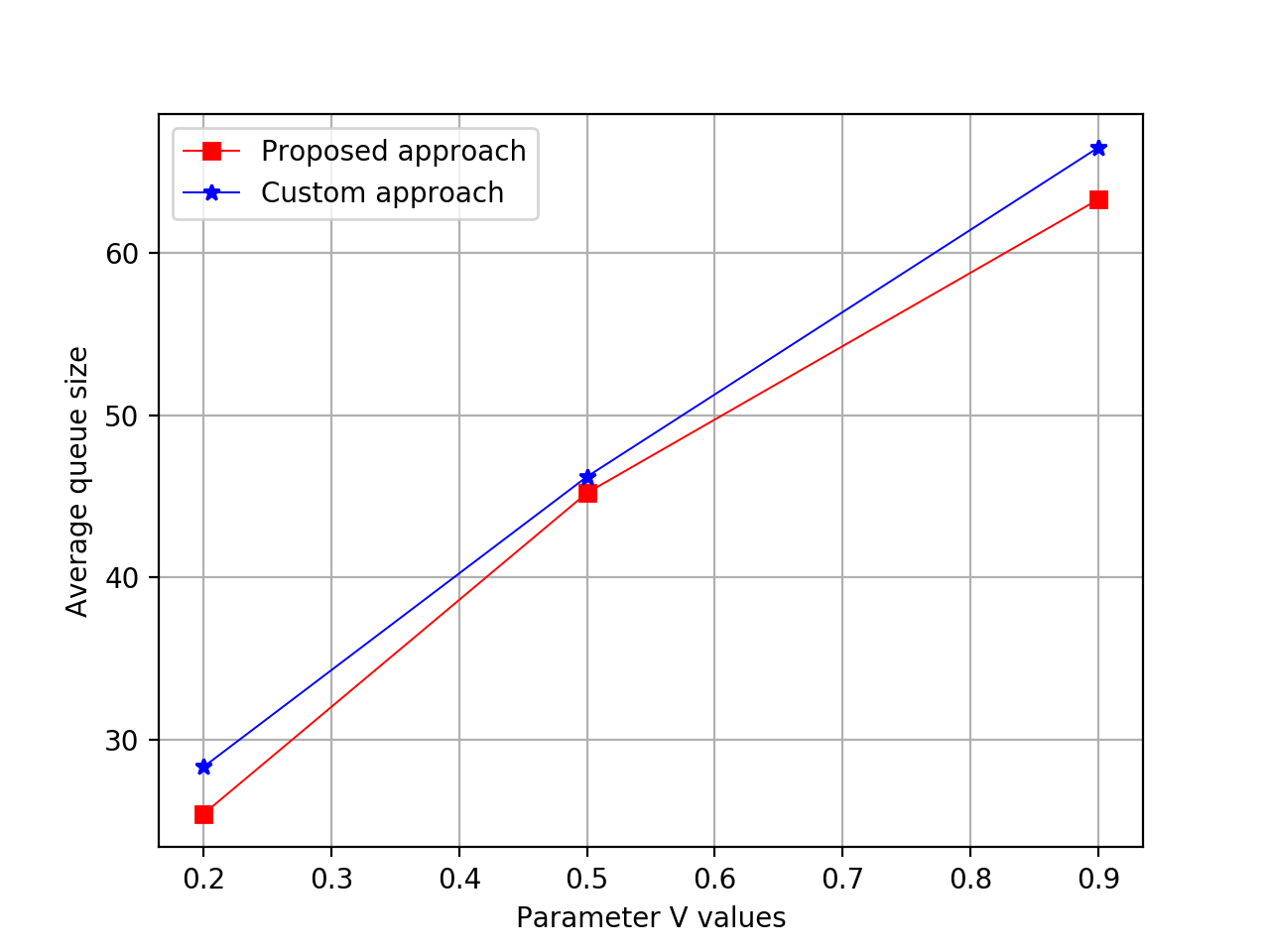}
		\caption{The impact of the parameter V on the queue size, in both approaches. V=\{0.2, 0.5, 0.9\}}.
		\label{fig:figure1v}
	\end{figure}
	
	On the requests arrival, the supervising entity check the available resources of each device through the ETSI MEC API. The response is the set of maximum capacities and the available resources of each device, also, checking for the containers queue state based on the equation (\ref{eq:queue_cont}). Once having the required information, the supervisor based solve the Lyapunov optimization in (\ref{eq:opt-prob}) to get the adequate policy $\pi_i$ (demand rate) to perform the tasks distribution and allocate the require resources.
	
		\begin{figure*}
		\centering
		\includegraphics[width=\linewidth]{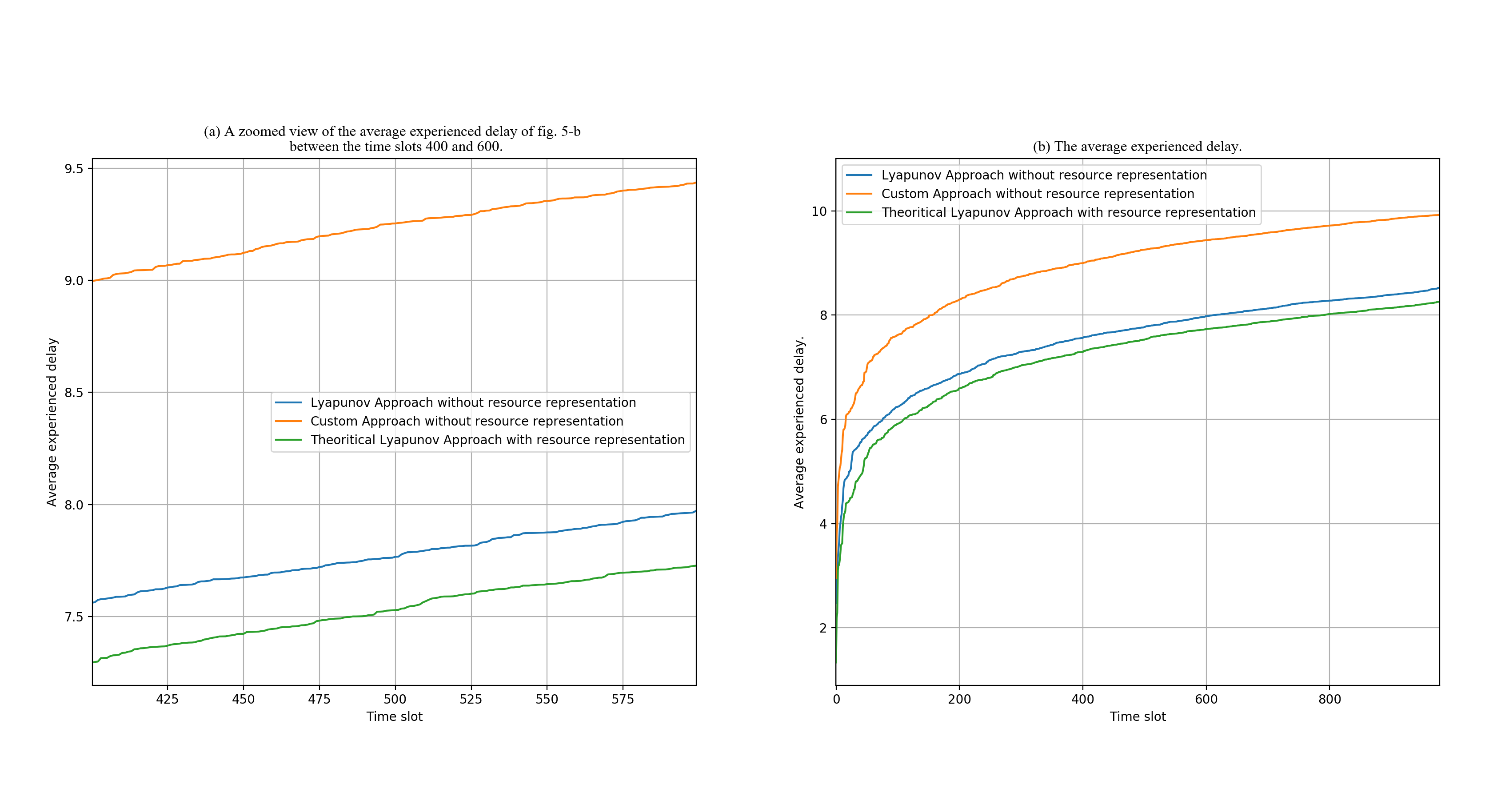}
		\caption{Delay evolution in time. (a) a zoomed view of the experienced delay between the time slots 400 and 600. (b) The average experienced delay (time slots)}
		\label{fig:delay}
	\end{figure*}
	
	We are interested in the experienced delay while processing requests on different fog nodes. We compared a custom approach, where no resource representation is considered, and our approach where the resource representation is considered. We enabled the resource representation using Lyapunov optimization, where requests are processed within lower delay and with a maximum number of requests processed as shown from the queue size in Fig. (\ref{fig:figure1v}).  The parameter $V$ ensures the trade-off between the experienced delay and the performance. The performance is characterized by the number of requests that a fog node can hold, in other words, it is the load of the node as described in (\ref{eq:alloc}). The more we have requests in the queue, the more delay increases. With a custom approach the queue size has more requests, which means that the number of requests to be processed is higher than when adopting a policy based on Lyapunov resource allocation policy. In addition, a small queue implies a lower delay, but in the case of a lower $V$ parameter, the users will not be interested in the delay, thus, more interested in what they are requesting. In this case, the queue size becomes bigger.
	
	Fig. \ref{fig:delay} shows the delay evolution in the different cases from the requests arrival, processing to response. Fig. (\ref{fig:delay}-a) shows the average experience delay in the time. In the custom approach adopted for resource allocation, the delay is high and the number of requests on pending goes high as well, and that because some devices are requested to perform tasks they are not capable to accomplish. Fig. (\ref{fig:delay}-b) is a zoomed view to reflect the details of the results obtained by applying Lyapunov approach, theoretically and applicably. The results show that the theoretical approach of the proposed Lyapunov approach give better results than results of applicative one by approximatively $\sim5\%$. The degradation of performance is due to the hardware losses. The results shown that the delays is decreased by approximately $\sim20\%$ in some cases.

	\section{Related Works}
	
	The work in \cite{cloud-resrep} proposed a model to represent resources such as computing, memory, storage, and communication within a cloud computing context. The proposed model is useful in order to characterize cloud applications, servers and even VMs.
	
	A holistic taxonomy on resource management at the edge of the network was proposed by \cite{taxonomy}. They present the terminology and the different types of architectures to characterize the state of the art in the edge computing paradigm. The paper discussed the resources management from four different perspectives, resource type, resource management objectives, resource location and resource use.
	
	Many papers proposed the utilization of containers at the edge of the network. In fact, using containers instead of the virtual machines could significantly reduce the overhead and response time. The authors in \cite{why-containers} propose an evaluation of the Docker containers technology as an Edge Computing platform. They evaluated the deployment, resources and service management, fault tolerance and finally the caching. They concluded that Docker containers provide fast deployment and good performance. This would make of containers a viable Edge Computing platform. Moreover, \cite{caas} within the containers technologies context at the Edge. They presented a novel architecture for tasks' management $-$selection and scheduling$-$ at the edge of using container-as-a-service (CoaaS). Finally, they proposed a container migration (internal and external) technique in real-time for minimizing the energy consumption.
	
	Lyapunov optimization was investigated by \cite{res-alloc-lyapunov} to solve the resource allocation problem. The authors investigate the adaptive resource allocation with Lyapunov optimization for time-varying wireless networks. They proposed a dynamic resource allocation (DRA) algorithm is to maintain the wireless network dynamics. Also, the authors investigate the tracking error between the results of DRA algorithm and the desired optimal resource allocation solution. Based on the results, they also develop an adaptive-compensation resource allocation (ACRA) algorithm, which iterates only once when the network state changes for saving the huge iteration overheads. Moreover, they proved that the ACRA algorithm asymptotically tracks the moving equilibrium point with no tracking errors.
	
	Authors in \cite{lyapunov} used Lyapunov optimization to reduce the energy consumption for reconfiguration in Cloud datacenters, while ensuring resources elasticity in order to ensure high availability of the platform. The servers capabilities were enhanced using the overclocking technique and Lyapunov optimization to ensure the the stability of the design and to minimize the energy cost.
	
	The game theory is a power mathematical framework widely used to solve the resource allocation problems. In \cite{res-alloc-game-th} the authors, consider a fog computing architecture, where a set of data service operators (DSO) controls the fog nodes (FN), providing data service requested by a set of data service subscribers (DSS). The authors proposed a joint optimization framework for all FN, DSO and DSS to achieve an optimal resource allocation scheme in a distributed fashion. The proposed approach consists of the pricing problem analyze for all DSOs and the resource allocation for DSSs. Moreover, a many-to-many matching game is applied to investigate the pairing problem between the DSOs and the FNs. Finally, within the same DSO, another many-to-many matching game was proposed to solve the FN-DSS pairing problem.
	
	Authors in \cite{rescont} build a new task-scheduling model within a containers architecture. They proposed a task-scheduling algorithm to ensure that the completion of tasks and the number of concurrent tasks for the fog node. Finally, a reallocation mechanism was proposed to reduce the latency in accordance with the characteristics of the containers.
	
	\section{Conclusion}\label{sec:conclusion}
	In this paper, we proposed a resource representation for fog computing through MEC APIs which allow exposing, at any time, information (i.e. CPU, memory, storage and networking) about the physical resources the fog devices. This information is used by the supervising entity of the fog to make the best decisions on distributing tasks to each node of the network. Moreover, we modeled the problem of latency minimization and proposed a corresponding Lyapunov-based optimization for resource allocation of the fog nodes. Simulation shows that the proposed approach, combining both the resource representation and the resource allocation optimization, allows minimizing the latency, thus, enhancing system performance.

\end{document}